\documentclass[aps,prl,twocolumn,superscriptaddress]{revtex4-1}
\usepackage{graphicx}
\usepackage{multirow}

\begin{document}

\title{Intercalation of Few-Layer Graphite Flakes with FeCl$_3$: Raman Determination of Fermi Level, Layer Decoupling and Stability}

\author{W. J. Zhao}
\author{P. H. Tan}
\author{J. Liu}
\affiliation{State Key Laboratory for Superlattices and
Microstructures, Institute of Semiconductors, Chinese Academy of
Sciences, Beijing 100083, China}
\author{A. C. Ferrari}
\affiliation{Department of Engineering, University of Cambridge, Cambridge CB3 0FA, UK}

\begin{abstract}
We use anhydrous ferric chloride (FeCl$_3$) to intercalate graphite flakes consisting of 2 to 4 graphene layers and to dope graphene monolayers. The intercalant, staging, stability and doping of the resulting intercalation compounds (ICs) are characterized by Raman scattering. The G peak of monolayer graphene heavily-doped by FeCl$_3$ upshifts to$\sim$1627cm$^{-1}$. 2-4 layer ICs have similar upshifts, and a Lorentzian lineshape for the 2D band, indicating that each layer behaves as a decoupled heavily doped monolayer. By performing Raman measurement at different excitation energies we show that, for a given doping level, the 2D peak can be suppressed by Pauli blocking for laser energy below the doping level. Thus, multi-wavelength Raman spectroscopy allows a direct evaluation of the Fermi level, complementary to that derived by Raman measurements at excitation energies higher than the doping level. We estimate a Fermi level shift of$\sim$0.9eV. These ICs are ideal test-beds for the physical and chemical properties of heavily-doped graphenes.
\end{abstract}

\maketitle

Graphite intercalation compounds (GICs) are host-guest systems\cite{Dresselhaus2002,Enoki2003}. They have been intensively studied because of novel features in their structures, electronic and optical properties\cite{Dresselhaus2002,Enoki2003,Caswell1978,Underhill1979,Cheng2009,Gruneis2009a,Gruneis2009b}. They are promising for electrodes, conductors, superconductors, catalysts, hydrogen storage materials, batteries, displays, polarizers\cite{Dresselhaus2002,Enoki2003,Caswell1978,Underhill1979,Pollak2010,Cheng2009,Gruneis2009a,Pollak2010}. Since the first synthesis of GICs in 1841\cite{Schaffautl1841}, hundreds of GICs have been produced with a variety of donor and acceptors intercalants\cite{Dresselhaus2002,Enoki2003}. GICs featuring unique stacking sequences (staging) have also been intensively studied\cite{Dresselhaus2002,Enoki2003,Caswell1978,Underhill1979}. In stage 1 GICs, each graphene layer is sandwiched by two intercalant layers\cite{Dresselhaus2002,Enoki2003,Caswell1978,Underhill1979,Gruneis2009a}. However, it is difficult to manipulate and process traditional GICs into nanoelectronic devices due to their thickness\cite{Dresselhaus2002,Enoki2003}. Graphene has great potential in nanoelectronics and optoelectronics\cite{Geim2007,Bonaccorso2010}. By intercalating graphite flakes just a few layers thick, one can combine the physical and chemical properties of GICs with those of few-layer graphenes (FLG) and open new opportunities for applications in nanoelectronics\cite{Das2008,Das2009,Chen2008,Yan2009,Mak2009,Efetov2010,Pachoud2010,Ye2010,Gunes2010,Hass2008}. There is also great interest on the transport properties of graphene at high carrier density, both for applications and fundamental physics\cite{Das2008,Das2009,Efetov2010,Pachoud2010,Ye2010}. By means of an electrolytic gate, Refs.\onlinecite{Das2008},\onlinecite{Das2009},\onlinecite{Pachoud2010} doped graphene up to$\sim$4.5$\times$10$^{13}$cm$^{-2}$. Ref.\onlinecite{Ye2010} used an ionic-liquid gate to achieve a carrier density higher than 10$^{14}$cm$^{-2}$. Ref.\onlinecite{Efetov2010} achieved 4$\times$10$^{14}$cm$^{-2}$ for hole and electron doping by means of solid polymer electrolyte. We note that in donor-type graphite intercalation compound, such as KC$_8$, the electron density can reach up to$\sim$4.8$\times$10$^{14}$cm$^{-2}$, corresponding to Fermi shift of$\sim$1.3eV\cite{Dresselhaus2002,Gruneis2009a,Gruneis2009b}. There is thus scope for using a similar approach to achieve graphene doped at levels higher than those reported in Refs.\onlinecite{Gruneis2009a,Das2008,Das2009,Efetov2010,Pachoud2010,Ye2010}.

Here, we use FeCl$_3$ to intercalate FLG and dope single layer graphene (SLG). Raman spectroscopy at several wavelengths shows the formation of acceptor-type Stage-1 GICs. We estimate a Fermi shift of$\sim$0.9eV, corresponding to a fractional charge transfer of$\sim$1/6.6 = 0.152 holes per carbon atom, i.e., a hole density of$\sim$5.8$\times$10$^{14}$cm$^{-2}$.
\begin{figure}[t]
\centerline{\includegraphics[width=90mm,clip]{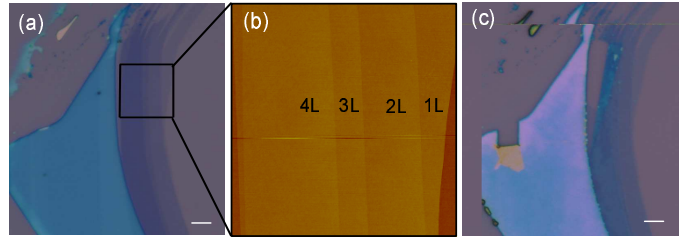}}
\caption{a) Optical image of a pristine 1-4L flake. b) AFM image of the region indicated by a solid square in (a). c) Optical image after FeCl$_3$ doping/intercalation. The scale bar in (a,c) is 4$\mu$m.} \label{fig1}
\end{figure}

Graphite flakes consisting of 1 to 4 layers (L) are obtained by micro-mechanical cleavage of natural graphite on a Si$+$300nm SiO$_2$ substrate\cite{Novoselov2005}. The number of layers is identified by optical contrast\cite{Casiraghi2007a,Blake2007} and atomic force microscopy (AFM)\cite{Novoselov2005}, Figs.1a,b. The Raman spectra are taken at room temperature using a Jobin-Yvon HR800 system with a$\sim$1.2cm$^{-1}$ spectral resolution. Intercalation is performed following the vapor transport method commonly used in GICs, as discussed, e.g., in Ref.\onlinecite{Dresselhaus2002}. The intended intercalant and the flakes are positioned in different zones in a glass ampoule, as shown in Fig.5b of Ref.\onlinecite{Dresselhaus2002}, then pumped to 1.5$\times$10$^{-4}$Torr. The reaction is carried out at 613K for six and thirty hours, for FLG and bulk graphite, respectively. A longer time is needed to reach stage 1 GIC for bulk graphite, due to the sample size, both in spatial extent and thickness. Fig.1c is an optical micrograph of a representative intercalated flake. The number of layers can still be identified, and their optical contrast is higher than prior to intercalation.

In GICs only a few layers thick it is difficult to apply X-ray diffraction techniques, unlike for bulk GICs staging determination\cite{Dresselhaus2002,Enoki2003,Caswell1978,Underhill1979}, because of the small flake thickness and the resulting substrate effects. Raman scattering was used to distinguish the intercalation and adsorption behavior of Bromine (Br$_2$), Iodine (I2), FeCl$_3$ and sulfuric acid\cite{Jung2009,Zhan2010,Zhao2010}. In principle, for stage-1 GICs, a single G peak is expected\cite{Dresselhaus2002,Enoki2003,Caswell1978,Underhill1979}. However, multiple G peaks were reported in recent works\cite{Zhan2010}.

The Raman spectrum of graphene consists of a set of distinct peaks. The G and D appear around 1580 and 1350cm$^{-1}$, respectively. The G peak corresponds to the E$_{2g}$ phonon at the Brillouin zone center. The D peak is due to the breathing modes of six-atom rings and requires a defect for its activation\cite{Ferrari2000a,Tuinstra1970,Thomsen2000}. It comes from TO phonons around the K point\cite{Ferrari2000a,Tuinstra1970,Ferrari2000b}, is active by double resonance (DR)\cite{Thomsen2000}, and strongly dispersive with excitation energy due to a Kohn Anomaly at K\cite{Piscanec2004}. DR can also happen intra-valley, i.e. connecting two points belonging to the same cone around K (or K$'$)\cite{Saito2002,Tan2002}. This gives the so-called D$'$ peak, which is at$\sim$1620cm$^{-1}$ in defected graphite measured at 514nm\cite{Tan2001}. The 2D peak is the second order of the D peak. This is a single peak in SLG, whereas it splits in four in bilayer graphene (BLG), reflecting the evolution of the band structure\cite{Ferrari2006}. Raman spectroscopy allows monitoring of doping, defects, strain, disorder, chemical modifications, edges, and relative orientation of the graphene layers\cite{Das2008,Das2009,Ferrari2000a,Piscanec2004,Ferrari2006,Pisana2007,Mohiuddin2009,Ferrari2007,Elias2009,Casiraghi2007b,Casiraghi2009,Yan2007,Yan2008,Graf2007}. Each Raman peak is characterized by its position, width, height, and area. The frequency-integrated area under each peak represents the probability of the whole process\cite{Basko2009}. We thus consider both area, A(2D)/A(G), and height, I(2D)/I(G), ratios.

The G peak position, Pos(G), has been widely used to identify staging\cite{Dresselhaus2002,Enoki2003,Caswell1978,Underhill1979}. In graphene, the shift of the Fermi energy has two major effects: (1) a change of the equilibrium lattice parameter with a consequent stiffening/softening of the phonons\cite{Das2008,Das2009,Pietronero1981}, and (2) the onset of effects beyond the adiabatic Born-Oppenheimer approximation, that modify the phonon dispersion close to the Kohn anomalies\cite{Das2008,Das2009,Piscanec2004,Pisana2007,Basko2009,Basko2008,Lazzeri2006}. (2) always results in a G upshift, both for electron and hole doping\cite{Piscanec2004,Pisana2007}, while (1) gives an upshift for $p$ doping and a downshift for $n$ doping. Thus for low doping levels, below$\sim$3$\times$10$^{13}$cm$^{-2}$, both $n$ and $p$ doping result in G peak upshifts\cite{Das2008,Das2009,Piscanec2004,Pisana2007,Lazzeri2006}, but for levels above$\sim$3$\times$10$^{13}$ cm$^{-2}$ the $n$ doped case reverses\cite{Lazzeri2006}. In fact, for $n$ doping$\sim$10$^{14}$cm$^{-2}$ the G shift would revert to zero\cite{Lazzeri2006}. Doping also significantly affects the 2D peak intensity. Ref. \onlinecite{Basko2009} reported that, for excitation energy well above the Fermi energy, this is linked to A(2D)/A(G) as:
\begin{equation}
\sqrt{\frac{A(G)}{A(2D)}}=\frac{0.26}{\gamma_{e-ph}}\left[\gamma_{e-ph}+|E_F|f(e^2/\varepsilon\nu_F)\right]
\label{eq1}
\end{equation}
where $\gamma_{e-ph}$, $E_F$, $e$, $\nu_F$ and $\varepsilon$ are the scattering rate due to the emission of phonons, Fermi energy, electron charge, electron velocity and dielectric constant\cite{Basko2009}.
\begin{figure}[t]
\centerline{\includegraphics[width=90mm,clip]{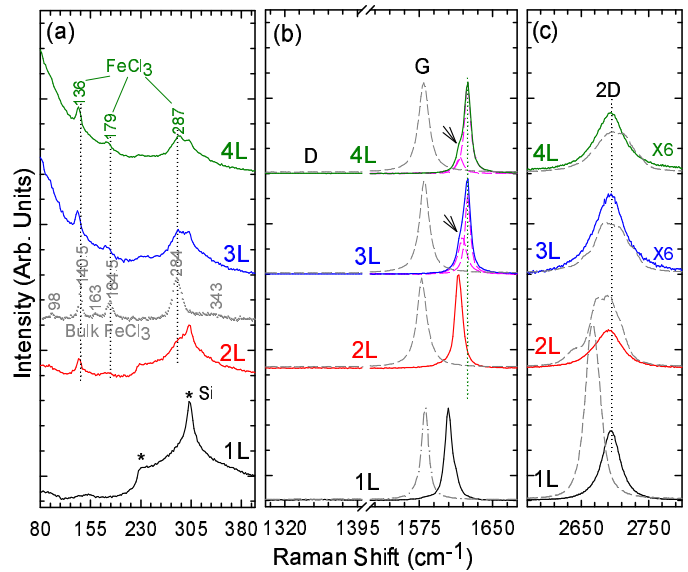}} \caption{
Raman spectra of pristine (dashed lines) and doped/intercalated (solid lines) 1-4L flakes, measured for 532nm excitation. a) Low frequency region. (b) D- and G-region. c) 2D-region. In (a) we also report the Raman spectrum of bulk FeCl$_3$ (dotted gray line) for comparison. Vertical dotted lines are guides to the eye} \label{fig2}
\end{figure}

FeCl$_3$ has eight Raman-active modes 4A$_g$+4E$_g$ (2A$_{1g}$+2A$_{2g}$+4E$_g$)\cite{Caswell1978}. Six Raman modes (3A$_g$+3E$_g$) were thus far experimentally measured\cite{Caswell1978}. When FeCl$_3$ was used as intercalant in stage 1 GICs, only four Raman modes (2A$_{1g}$+2E$_g$) were observed\cite{Caswell1978}. The other two A$_g$ and E$_g$ modes at$\sim$164 and 354cm$^{-1}$ are probably too weak to be observed in GICs. Indeed, even in crystal FeCl$_3$, those modes are very weak\cite{Caswell1978}. Fig. 2a shows that, after doping by FeCl$_3$, three Raman modes from FeCl$_3$ are observed in the low frequency region: A$_{1g}$$\sim$136cm$^{-1}$, E$_g$$\sim$179cm$^{-1}$ and A$_{1g}$$\sim$287cm$^{-1}$. We cannot detect the other E$_g$ mode$\sim$93cm$^{-1}$, since this is too weak to be distinguished from the background. These peaks positions are very close to those previously observed for FeCl$_3$ intercalated Stage-1 GICs\cite{Dresselhaus2002,Enoki2003,Caswell1978}, but differ from bulk crystalline FeCl$_3$, whose spectrum is also shown in Fig.2a for comparison: the A$_{1g}$ (at 136cm$^{-1}$) and E$_g$ mode (at 179cm$^{-1}$) have a$\sim$4cm$^{-1}$ downshift, while the A$_{1g}$ (at 287cm$^{-1}$) upshifts$\sim$3cm$^{-1}$. In bulk FeCl$_3$, the iron layer is sandwiched by two chlorine layers, as shown in Fig.1 of Ref.\onlinecite{Caswell1978}. When intercalation happens, Cl atoms simultaneously occupy preferred sites associated with the graphene lattice, which results in the loss of the Cl atoms long-range two dimensional order, because their in-plane structure is incommensurate with the graphene host lattice\cite{Dresselhaus2002,Caswell1978}. Fe atoms however retain the long-range order as in crystal FeCl$_3$\cite{Dresselhaus2002,Caswell1978}. This results in a$\sim$3$^o$ relative rotation of the Cl layers above and below the Fe layer, and a difference of the Raman modes of intercalated FeCl$_3$ compared to bulkFeCl$_3$\cite{Dresselhaus2002,Caswell1978}. FeCl$_3$ modes are not observed in the FeCl$_3$-doped SLG due to the very low density of FeCl$_3$, compared to fully- FeCl$_3$-intercalated GICs, due to de-adsorption.
\begin{figure}[t]
\centerline{\includegraphics[width=90mm]{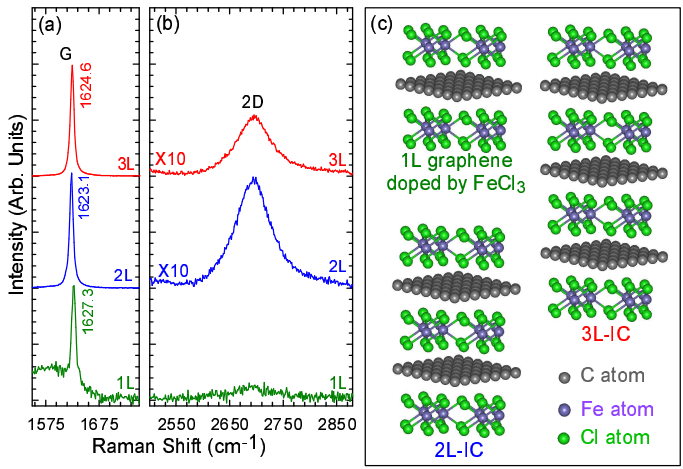}}
\caption{a) G and b) 2D band of Stage-1 flakes with 2/3L, and doped SLG, measured at 532nm for samples kept in the sealed quartz tube used for intercalation/doping. c) Schematic diagrams of FeCl$_3$ doped/intercalated 1-3L flakes\cite{Dresselhaus2002}} \label{fig3}
\end{figure}

Figure 2 plots the Raman spectra of 1-4L flakes measured at 532nm before (dashed lines) and after (solid lines) FeCl$_3$ intercalation. The pristine samples have the characteristic features of mono- and multi-layer graphene\cite{Ferrari2006,Ferrari2007}. The Pos(G) shift of doped/intercalated samples compared to pristine ones in Fig.2b is a signature of doping. The blueshift is smaller in SLG compared to FLG. This indicates fewer FeCl$_3$ molecules on SLG relative to 2-4L flakes. The SLG 2D band in Fig.2c upshifts$\sim$28cm$^{-1}$, typical of hole-doping\cite{Das2008,Das2009}, while I(2D)/I(G) and A(2D)/A(G) decrease$\sim$61\% and 53\% relative to those prior to doping. From Refs.\onlinecite{Das2009,Basko2009}, we estimate the Fermi shift for SLG to be$\sim$0.4eV.

The 2D lineshape for the 2-4L flakes after FeCl$_3$ intercalation changes significantly, as shown in Fig.2c, from multiple peaks to a single Lorentzian. This is an indication of electronic decoupling of the layers\cite{Ferrari2006,Ferrari2007}. Note that the residual presence of any non-intercalated BLG or FLG would give residual multiple 2D bands. In pristine Bernal-stacked graphite, the interlayer distance is 3.35$\AA$\cite{Dresselhaus2002,Enoki2003}. When FeCl$_3$ molecules are intercalated, the distance increases to 9.37$\AA$\cite{Dresselhaus2002,Enoki2003,Caswell1978,Underhill1979}. As a result, the interlayer interaction significantly decreases\cite{Dresselhaus2002,Enoki2003,Gruneis2009a}. Therefore, the single Lorentzian 2D peak indicates SLG between two intercalant layers, each hole-doped.
\begin{figure}[t]
\centerline{\includegraphics[width=90mm,clip]{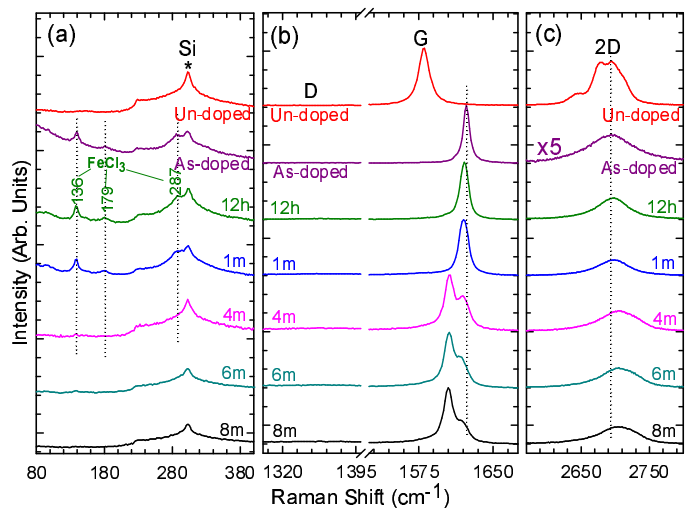}}
\caption{Raman spectra for 532nm excitation energy of as-prepared 2L-GICs and after 12 hours, and 1, 4, 6, 8 months: (a) Low frequency region; (b) D and G band region; (c) 2D band region. A pristine BLG is also included as reference. Vertical dotted lines are guide to the eyes.} \label{fig4}
\end{figure}

As shown in Fig.2b, the G bands of SLG and BLG are$\sim$1605 and$\sim$1615cm$^{-1}$, moving to$\sim$1625cm$^{-1}$ for 3L and 4L. The latter value is close to that previously reported in FeCl$_3$ Stage-1 GICs ($\sim$1626cm$^{-1}$)\cite{Dresselhaus2002,Enoki2003,Caswell1978,Underhill1979}. For 3L and 4L, an additional sideband appears$\sim$1618cm$^{-1}$, with almost equal width to the main peak. One might assign it as the D$'$ peak activated by defects. However, this is not the case for our samples since the D peak is unobservable for all the layers, before and after doping/intercalation, Fig.2b. This band is thus another G peak, resulting from non-uniform intercalation, due to de-adsorption1 of FeCl$_3$ when the flakes are exposed to air. This would imply that the shoulder is due to the upper or lower layers. The thicker the flakes, the less the top and bottom layers will contribute to the overall intensity of the measured Raman spectrum. Indeed, 4L flakes have weaker relative intensity of this shoulder compared to the main G peak, than 3L flakes. No such shoulder is expected in SLG, where the lower Pos(G) is an indication of lower coverage. We note that no shoulder is observed for the doped/intercalated BLG, indicative of homogenous doping. Let us consider the upper and lower layers of intercalated 2-4L flakes. If FeCl$_3$ is only present on one side of these layers, we expect the amount of charge transfer to reach at most that of Stage-2 GICs. In this case Pos(G) can only shift to$\sim$1612cm$^{-1}$, i.e. Pos(G) of FeCl$_3$-intercalated Stage-2 GICs\cite{Dresselhaus2002,Enoki2003,Caswell1978,Underhill1979}. Pos(G) of our FeCl$_3$-doped SLG is very close to that reported in Ref.\onlinecite{Zhan2010}, where double G peaks, at$\sim$1612 and$\sim$1623cm$^{-1}$ were also observed for 3 and 4L samples. Ref.\onlinecite{Zhan2010} explained this by arguing that FeCl$_3$ did not adsorb on the top and bottom of their flakes. We note that Pos(G) of our intercalated BLG after air exposure ($\sim$1615cm$^{-1}$) and of the lower energy G band in 3-4L FeCl$_3$-intercalated flakes are higher than previously reported for FeCl$_3$ intercalated Stage 2 GICs\cite{Dresselhaus2002,Enoki2003,Caswell1978,Underhill1979}. This means that the upper and lower surfaces of our FeCl$_3$-intercalated 2-4L are doped, but not with full coverage, due to FeCl$_3$ de-adsorption, unlike the interior layers. The lower Pos(G) in FeCl$_3$-intercalated BLG compared to 3 and 4L is expected, since a BLG has no fully enclosed SLG, thus de-adsorption effects are expected to be stronger.
\begin{figure}[t]
\centerline{\includegraphics[width=80mm,clip]{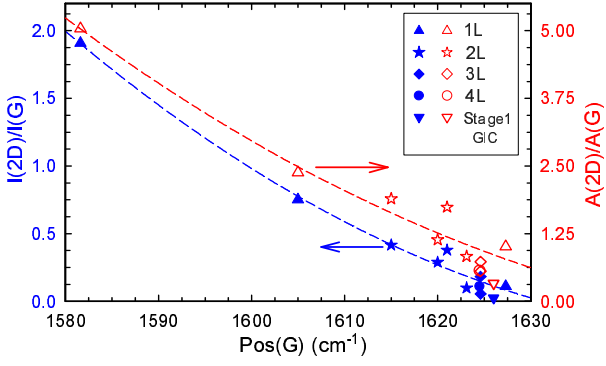}}
\caption{I(2D)/I(G) and A(2D)/A(G) as a function of Pos(G). The blue and red dashed lines are guide to the eyes} \label{fig5}
\end{figure}

If the samples are quickly removed from the sealed tube and cooled to room temperature in air, their Raman spectra are very similar to those previously reported for Stage-1 GICs\cite{Dresselhaus2002,Enoki2003,Caswell1978,Underhill1979}. Fig.3a shows Pos(G) of intercalated 2 and 3L$\sim$1623 and 1625cm$^{-1}$, very close to that of Stage-1 GICs\cite{Dresselhaus2002,Enoki2003,Caswell1978,Underhill1979}. However, when measuring in-situ the SLG Raman spectrum, i.e. keeping the sample sealed in a quartz tube after cooling at room temperature following a 30 minutes doping, Pos(G) reaches$\sim$1627cm$^{-1}$, indicative of heavy doping. This is slightly higher than previously reported for Stage-1 GICs\cite{Dresselhaus2002,Enoki2003,Caswell1978,Underhill1979}, and much higher than Stage-2 GICs\cite{Dresselhaus2002,Enoki2003,Caswell1978,Underhill1979}. This is thus evidence of doping on both top and bottom surfaces of our SLG. When this sample is removed from the tube and exposed to air, Pos(G) goes to$\sim$1605cm$^{-1}$, similar to Fig.2, due to desorption. Figs.3a,b show that Pos(G) for BLG is lower than in 3L, and I(2D)/I(G) for BLG stronger than 3L. This suggests de-intercalation in BLG flakes.

To further study the de-adsorption/de-intercalation, Figs.4a-c report the BLG Raman spectra as a function of time for a period of up to 8 months. Pos(G) starts at$\sim$1623cm$^{-1}$, and can be fitted with a single Lorentzian with FWHM(G)$\sim$8cm$^{-1}$, indicating uniform doping\cite{Das2008,Das2009,Pisana2007,Casiraghi2007b}. After twelve hours, Pos(G) decreases to$\sim$1621cm$^{-1}$, and FWHM(G) increases to$\sim$10cm$^{-1}$. After one month, Pos(G)$\sim$1620cm$^{-1}$ and FWHM(G)$\sim$12cm$^{-1}$. The spectrum evolution indicates that FeCl$_3$-intercalated flakes are relatively stable in air at room temperature for up to one month. The intercalant Raman modes change little within one month. However, after four months, they are not seen anymore, while G and 2D acquire a multiple peak profile. This confirms that a significant reduction in doping occurs after one month due to de-adsorption and de-intercalation. The multiple G peaks after 4 months may result from different de-adsorption rates on different surfaces.

Figure 5 plots I(2D)/I(G) and A(2D)/A(G) as a function of Pos(G). With increasing Pos(G), i.e. increasing doping, I(2D)/I(G) and A(2D)/A(G) both decrease. I(2D)/I(G) and A(2D)/A(G) of heavily doped SLG are close to those of almost fully-doped 2 and 3L, further confirming double-surface doping.

We now consider the dependence of I(2D)/I(G) and A(2G)/A(G) on doping and excitation wavelength. We use a FeCl$_3$-intercalated Stage-1 GIC as an example to show how to probe the Fermi level by multi-wavelength Raman spectroscopy, since in this case the Fermi level is well know by independent characterizations\cite{Dresselhaus2002}. Fig.6a plots the Raman spectra measured at 488, 514, 561, 593 and 633nm. These are similar to those in Fig.3. For all lasers Pos(G)$\sim$1626cm$^{-1}$,FWHM(G)$\sim$7cm$^{-1}$, in good agreement with that previously reported for FeCl$_3$-intercalated Stage-1 GICs\cite{Dresselhaus2002,Enoki2003,Caswell1978,Underhill1979}. At 633nm, the 2D peak is almost unobservable, similarly to what reported in Ref.\onlinecite{Jung2009} for SLGs doped by bromine. However, increasing the excitation energy from 2.09eV (593nm) to 2.54eV (488nm), the 2D peak appears. A(2D)/A(G) and I(2D)/I(G) are plotted as a function of excitation energy in Fig.6b. The trend of these intensity ratios can be understood considering the full resonant Raman scattering process for the 2D band\cite{Basko2009,Basko2008}. Fig.6c plots the doped SLG band structure. For a given laser energy, to activate the 2D peak an electron-hole pair must be excited in process: a$\rightarrow$b, and recombined in process c$\rightarrow$d. These transitions differ by the 2D peak energy:
\begin{equation}
E_T=E_L-\hbar\omega_{2D}
\label{eq2}
\end{equation}
There are three cases 1) When $E_L$ and $E_T$ are both larger than 2$E_F$, the 2D band can be always observed; 2) When $E_L$ is larger than 2 $E_F$, but $E_T$ is smaller than 2$E_F$, process c$\rightarrow$d is forbidden due to Pauli blocking; 3) when both $E_L$ and $E_T$ are smaller than 2$E_F$, both processes a$\rightarrow$b and c$\rightarrow$d are forbidden. Therefore, only when $E_T$$>$2$E_F$, i.e., ($E_L$-$\hbar\omega_{2D}$)/2$>$$E_F$, the 2D band is observable. Thus, the absence of the 2D band in the Raman spectra of FeCl$_3$-intercalated Stage-1 GICs, and FeCl$_3$ and Br$_2$ heavily-doped SLGs\cite{Jung2009} at 1.96eV (632.8nm) indicates that their $E_F$ is larger than 0.81eV. When ($E_L$-$\hbar\omega_{2D}$)/2$>$$E_F$, both I(2D)/I(G) and A(2D)/A(G) should increase. Thus, the sharp intensity increase when moving from 2.21eV (561nm) to 2.09eV (593nm), implies that $E_T$ corresponding to E$_L$=2.09eV is close to 2$E_F$. Then, $E_F\sim$0.88eV, close to that measured in Stage-1 GICs by electron energy loss spectroscopy\cite{Dresselhaus2002}.
\begin{figure}[t]
\centerline{\includegraphics[width=70mm,clip]{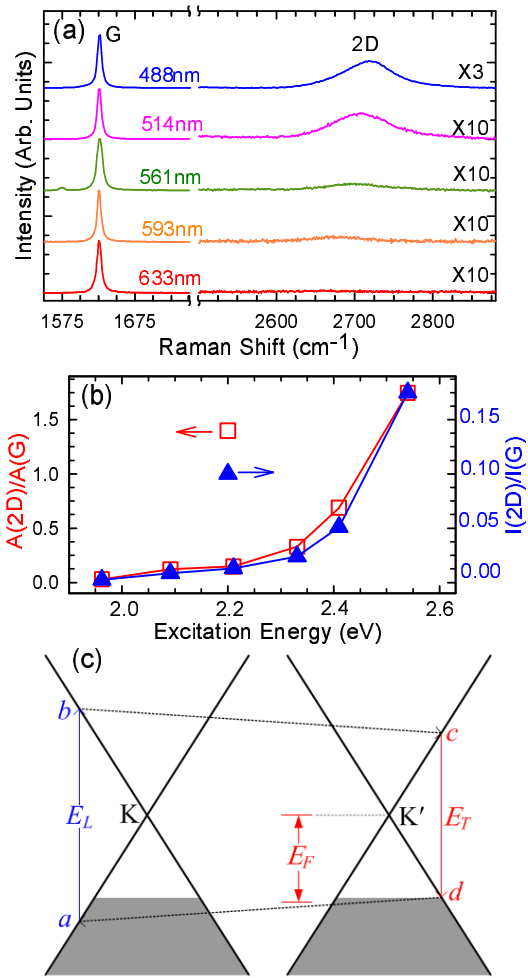}}
\caption{a) Raman spectra of stage-1 GIC measured at 488, 514 532, 561, 593 and 633nm, normalized to have the same G band intensity. b) A(2D)/A(G) and I(2D)/I(G) as a function of excitation energy. c) Schematic diagram of SLG band structure and fully resonant Raman processes for the 2D band: a$\rightarrow$b, photon absorption; c$\rightarrow$d, electron-hole recombination; b$\rightarrow$c and d$\rightarrow$a, phonon emission.} \label{fig6}
\end{figure}

$E_F$ can be also estimated from A(G)/A(2D) at 514nm, by using Eq.1. The numerical values of $f(e^2/\varepsilon\nu_F)$ can be taken from Fig.3 in Ref.\onlinecite{Basko2009}, and $\gamma_{e-ph}$=21meV can be extracted from the hole-doping data of Ref.\onlinecite{Das2009} measured at 514nm. While $\varepsilon$ is not available for intercalated FeCl$_3$, we can estimate it from that measured for FeCl$_3$ in aqueous solutions\cite{S1}. This gives $f$=0.0908. Then, inserting A(2D)/A(G)=0.688 in Eq.1, gives E$_F\sim$0.84eV, very close to that derived by the intensity transition as a function of excitation energy.

In conclusion, graphite flakes consisting of a few graphene layers can be doped by adsorption and intercalation of FeCl$_3$. This results in each of the layers behaving as a hole-doped SLG. These are stable up to one month after air exposure. The variation of the 2D intensity relative to the G peak with excitation energy allows one to estimate the Fermi energy. We get $E_F$$\sim$0.9eV, corresponding to a fractional charge transfer of$\sim$1/6.6=0.152 holes per carbon, i.e.$\sim$5.8$\times$10$^{14}$cm$^{-2}$, larger than the 4$\times$10$^{14}$cm$^{-2}$ recently reported by employing a solid polymer electrolyte gate\cite{Efetov2010}.

\textbf{Acknowledgments}. This work was supported by the National Natural Science Foundation of China under Grant Nos. 10934007 and 10874177, and the special funds for the Major State Basic Research under Contract No. 2009CB929300 of China. ACF acknowledges funding from ERC grant NANOPOTS, EPSRC grant EP/G042357/1, Royal Society Wolfson Research Merit Award, EU grants RODIN and GENIUS.


\begin{thebibliography}{52}
\bibitem{Dresselhaus2002}M.S.Dresselhaus,G.Dresselhaus,Adv. Phys. \textbf{51},1 (2002)
\bibitem{Enoki2003}T. Enoki, M. Suzuki, M. Endo, \textit{Graphite intercalation compounds and applications}. Oxford, (2003)
\bibitem{Caswell1978}N. Caswell, S. A. Solin, Solid State Comm. \textbf{27} 961 (1978)
\bibitem{Underhill1979}C. Underhill {\it et al.}, Solid State Comm. \textbf{29}, 769(1979).
\bibitem{Cheng2009}H. S. Cheng {\it et al.}, J. Am. Chem. Soc. \textbf{131}, 17732(2009).
\bibitem{Gruneis2009a}A. Gruneis {\it et al.}, Phys. Rev. B \textbf{79}, 205106 (2009).
\bibitem{Gruneis2009b}A. Gruneis {\it et al.}, Phys. Rev. B \textbf{80}, 075431 (2009).
\bibitem{Pollak2010}E. Pollak {\it et al.}, Nano Lett. \textbf{10}, 3386 (2010).
\bibitem{Schaffautl1841}P. Schaffautl, J. prakt. Chem. \textbf{21}, 155 (1841).
\bibitem{Geim2007}A. K. Geim, K. S. Novoselov, Nat. Mater. \textbf{6}, 183 (2007).
\bibitem{Bonaccorso2010}F. Bonaccorso {\it et al.}, Nat. Photonics \textbf{4}, 611 (2010).
\bibitem{Das2008}A. Das {\it et al.}, Nat. Nanotechnol. \textbf{3}, 210 (2008).
\bibitem{Das2009}A. Das {\it et al.}, Phys. Rev. B \textbf{79}, 155417 (2009).
\bibitem{Chen2008}J. H. Chen {\it et al.}, Nat. Nanotechnol. \textbf{3}, 206 (2008).
\bibitem{Yan2009}J. Yan {\it et al.}, Phys. Rev. B \textbf{80}, 241417 (2009).
\bibitem{Mak2009} K. F. Mak {\it et al.}, Phys. Rev. Lett. \textbf{102}, 256405 (2009).
\bibitem{Efetov2010}D. K. Efetov, and P. Kim, ArXiv:1009.2988v1 2010.
\bibitem{Pachoud2010}A. Pachoud {\it et al.}, ArXiv:1009.3367v1 2010.
\bibitem{Ye2010}J. T. Ye {\it et al.}, ArXiv:1010.4679v1 2010.
\bibitem{Gunes2010}F. Gunes {\it et al.}, Acs Nano \textbf{4}, 4595 (2010).
\bibitem{Hass2008}J. Hass {\it et al.}, Phys. Rev. Lett. \textbf{100}, 125504 (2008).
\bibitem{Novoselov2005} K. S. Novoselov {\it et al.}, PNAS \textbf{102}, 10451 (2005).
\bibitem{Casiraghi2007a} C. Casiraghi {\it et al.}, Nano Lett. \textbf{7}, 2711 (2007).
\bibitem{Blake2007}P. Blake {\it et al.}, Appl. Phys. Lett. \textbf{91}, 063124 (2007).
\bibitem{Jung2009}N. Jung {\it et al.}, Nano Lett. \textbf{9}, 4133 (2009).
\bibitem{Zhan2010}D. Zhan {\it et al.}, Adv. Funct. Mater. \textbf{20}, 3504 (2010).
\bibitem{Zhao2010}W. J. Zhao {\it et al.}, Phys. Rev. B 2010, to be published.
\bibitem{Ferrari2000a}A.C.Ferrari,J.Robertson,Phys. Rev. B \textbf{61},14095 (2000)
\bibitem{Tuinstra1970}F. Tuinstra,J. L. Koenig, J. Chem. Phys. \textbf{53}, 1126 (1970).
\bibitem{Thomsen2000}C. Thomsen,S. Reich,Phys. Rev. Lett. \textbf{85},5214 (2000)
\bibitem{Ferrari2000b}A. C. Ferrari {\it et al.}, Phys. Rev. B \textbf{62}, 11089 (2000).
\bibitem{Piscanec2004}S. Piscanec {\it et al.}, Phys. Rev. Lett. \textbf{93}, 185503 (2004).
\bibitem{Saito2002}R. Saito {\it et al.}, Phys. Rev. Lett. \textbf{88}, 027401 (2002).
\bibitem{Tan2002}P. H. Tan {\it et al.}, Phys. Rev. B \textbf{66}, 245410 (2002).
\bibitem{Tan2001}P. H. Tan {\it et al.}, Phys. Rev. B \textbf{64}, 214301 (2001).
\bibitem{Ferrari2006}A. C. Ferrari {\it et al.}, Phys. Rev. Lett. \textbf{97}, 187401 (2006).
\bibitem{Pisana2007}S. Pisana {\it et al.}, Nat. Mater. \textbf{6}, 198 (2007).
\bibitem{Mohiuddin2009}T.M.G.Mohiuddin {\it et al.},Phys. Rev. B \textbf{79}, 205433 (2009)
\bibitem{Ferrari2007}A. C. Ferrari, Solid State Commun. \textbf{143}, 47 (2007).
\bibitem{Elias2009}D. C. Elias {\it et al.}, Science \textbf{323}, 610 (2009).
\bibitem{Casiraghi2007b}C. Casiraghi {\it et al.}, Appl. Phys. Lett. \textbf{91}, 233108 (2007).
\bibitem{Casiraghi2009}C. Casiraghi {\it et al.}, Nano Lett. \textbf{9}, 1433 (2009).
\bibitem{Yan2007}J. Yan {\it et al.}, Phys. Rev. Lett. \textbf{98}, 166802 (2007).
\bibitem{Yan2008}J. Yan {\it et al.}, Phys. Rev. Lett. \textbf{101}, 136804 (2008).
\bibitem{Graf2007}D. Graf {\it et al.}, Nano Lett. \textbf{7}, 238 (2007).
\bibitem{Basko2009}D. M. Basko, S. Piscanec, A. C. Ferrari, Phys. Rev. B \textbf{80}, 165413 (2009).
\bibitem{Pietronero1981}L.Pietronero,S.Strassler, Phys. Rev. Lett.\textbf{47}, 593 (1981)
\bibitem{Basko2008}D. M. Basko, Phys. Rev. B \textbf{78}, 125418 (2008).
\bibitem{Lazzeri2006}M.Lazzeri,F.Mauri,Phys. Rev. Lett. \textbf{97},266407 (2006)
\bibitem{Latil2007}S. Latil et al. Phys. Rev. B \textbf{76}, 201402 (2007).
\bibitem{S1} N. A. El-Shistawi, M. A. Hamada and E. A. Gomaa, Chemistry \textbf{18}, 5 (2009).
\end{thebibliography}
\end{document}